\documentclass[twocolumn,preprintnumbers,amsmath,amssymb,aip]{revtex4}

\usepackage{graphicx}
\usepackage{dcolumn}
\usepackage{bm}
\usepackage{color}
\usepackage{setspace}

\begin{document}


\title{Efficient Edelstein effects in one-atom-layer Tl-Pb compound}

\author{Y. Shiomi$^{\, 1}$}
\altaffiliation[Present address: ]{Department of Applied Physics and Quantum-Phase Electronics Center (QPEC), University of Tokyo, Hongo, Tokyo 113-8656, Japan, and RIKEN Center for Emergent Matter Science (CEMS), Wako 351-0198, Japan}
\author{K. T. Yamamoto$^{\, 1}$}
\author{R. Nakanishi$^{\, 2}$}
\author{T. Nakamura$^{\, 2}$}
\author{S. Ichinokura$^{\, 2}$}
\author{R. Akiyama$^{\, 2}$}
\author{S. Hasegawa$^{\, 2}$}
\author{E. Saitoh$^{\, 1,3,4,5}$}
\altaffiliation[Present address: ]{Department of Applied Physics, University of Tokyo, Hongo, Tokyo 113-8656, Japan}

\affiliation{$^{1}$
Institute for Materials Research, Tohoku University, Sendai 980-8577, Japan}
\affiliation{$^{2}$
Department of Physics, University of Tokyo, Tokyo 113-0033, Japan}
\affiliation{$^{3}$
Advanced Institute for Materials Research, Tohoku University, Sendai 980-8577, Japan
}
\affiliation{$^{4}$
Advanced Science Research Center, Japan Atomic Energy Agency, Tokai 319-1195, Japan
}
\affiliation{$^{5}$
Center for Spintronics Research Network, Tohoku University, Sendai 980-8577, Japan}

\date{\today}

\begin{abstract}
We have investigated direct and inverse Edelstein effects in a one-atom-layer Tl-Pb compound with a large Rashba-type spin splitting. In spin pumping experiments at room temperature, spin-to-charge conversion voltage due to the inverse Edelstein effect is clearly observed in Py/Cu/Tl-Pb trilayer samples. To confirm efficient spin-charge interconversion in Tl-Pb compounds, the direct Edelstein effect is also studied in the same trilayer samples by measuring modulation of the effective magnetization damping in the Py layer via the charge-to-spin conversion in the Tl-Pb layer. Using both the results of direct and inverse Edelstein effects, the Edelstein length is estimated to be $\sim 0.1$ nm for Tl-Pb compounds.     
\end{abstract}

\maketitle

Two-dimensional electronic states have recently attracted much attention in the spintronics field \cite{weihan}. Space-inversion asymmetry leads to spin splitting of two-dimensional electronic states due to the Rashba-type spin-orbit interaction \cite{kush2006, kush2007, kush2008}. The spin split bands are characterized by Fermi contours with helical locking of spin with momentum. The helical spin texture accompanies nonzero spin accumulation along an inplane direction transverse to the direction of applied electric currents, which is known as an Edelstein effect \cite{Edelstein}. As a spin-to-charge conversion effect in the two-dimensional systems, an inverse effect of the Edelstein effect, the inverse Edelstein effect \cite{sanchez}, has been studied by using the spin pumping technique on the surface of topological insulators \cite{shiomi, deorani, jameli, Sn, yamamoto, HWang, smb6} and the interface of bilayer materials \cite{sanchez, karube, tsai, whan}. Since the Edelstein effects on two-dimensional Rashba states have potentially better efficiencies of spin-charge interconversion than the inverse spin Hall effect observed in conventional heavy metals \cite{ralph, fan}, material search for the efficient Edelstein effects could be important for future spintronics applications.
\par
 
Atomic-layer metal films on Si (silicon) substrates have been intensively studied in the field of surface science for more than 50 years \cite{lander}. Because of the surface reconstruction on Si, surface superstructures can possess conducting surface states, which are inherently two dimensional and decoupled from the bulk. Although conductivity of ultrathin films is usually suppressed by reducing the film thickness, monolayers of In- and Pb-induced surface superstructures on Si (111) have good conductivity and even show superconductivity at low temperatures ($1$-$3$ K) \cite{zhang, qin, uchihashi, yamada, uchihash2, yoshizawa, noffsinger, zhao, brun}. 
\par

Recently, it was reported that alloying one monolayer of Tl with one-third monolayer of Pb results in a one-atom-layer Tl-Pb compound on Si (111) that exhibits both two-dimensional superconductivity ($2.25$ K) and Rashba-type spin splitting \cite{matetskiy}. Angle-resolved photoelectron spectroscopy showed that the magnitude of the Rashba-type spin splitting reaches $\sim$250 meV \cite{matetskiy}, which is even larger than that in one-third monolayer of Bi on Ag surface alloys ($\sim$200 meV) \cite{ast}. Although giant Rashba-type spin splitting has been also observed in Bi \cite{gierz, sakamoto, frantzeskakis} and Tl \cite{sakamoto2, azpiroz, sakamoto3, stolwijk, stolwijk2} monolayers on Si (111) surfaces, these systems are non-metallic, not suitable for spintronics applications. Hence, the Tl-Pb compounds on Si are rare atomic-layer materials promising for spin-charge interconversion due to the Edelstein effects.
\par

In this manuscript, we report spin-charge interconversion effects in Tl-Pb compounds. By spin pumping from a ferromagnetic permalloy (Py) layer to the Tl-Pb layer, large Lorentz-type voltage peaks were observed at ferromagnetic resonance (FMR) of Py, while small anti-symmetric voltage signals due to ferromagnetic transport was observed only in a control sample without the Tl-Pb compound. The dominant Lorentz-type voltage signal induced by spin pumping into the Tl-Pb layer strongly indicates that the inverse Edelstein effect emerges on the Tl-Pb compound. Furthermore, to evaluate the efficiency of the Edelstein effects in the Tl-Pb compound reliably, we also measured the magnetization-damping modulation of the Py layer induced by spin transfer torque originating from the Edelstein effect on the Tl-Pb layer. From the experimental results of the Edelstein effect and the inverse Edelstein effect, the Edelstein length \cite{sanchez}, which is a measure of the spin-charge conversion efficiency, was estimated to be $\sim 0.1$ nm for Tl-Pb; this magnitude is comparable to that reported in Py/topological-insulator samples, {\it e.g.}  Py/Bi$_2$Se$_3$ \cite{deorani} and Py/Cu/Sn-Bi$_2$Te$_2$Se \cite{yamamoto}.
\par

\begin{figure}[t]
\begin{center}
\includegraphics[width=8cm]{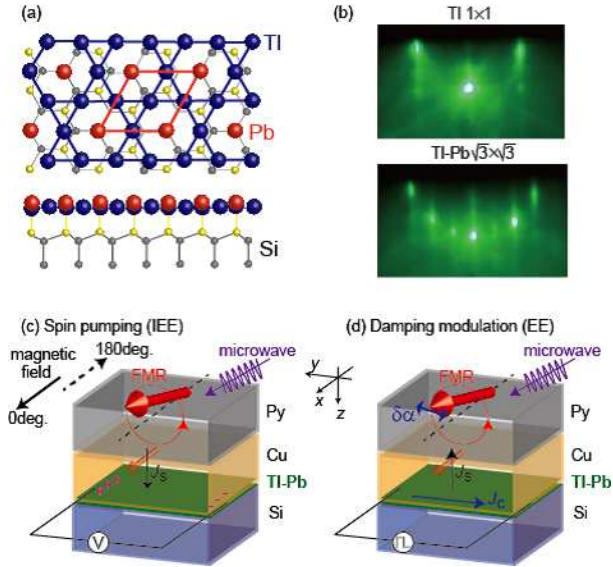}
\caption{(Color online.) (a) Atomic structure of the Tl-Pb compound grown on a Si(111) substrate. Tl and Pb atoms are shown by blue and red circles, respectively. Si atoms are also shown by yellow and gray circles. The $\sqrt{3} \times \sqrt{3}$ unit cell is outlined by a red frame. (b) In situ RHEED patterns during the growth process of a Tl-Pb compound. (c) Schematic illustration of the inverse Edelstein effect (IEE) induced by spin pumping into the Tl-Pb layer. (d) Schematic illustration of modulation of the effective damping constant via spin transfer through the Edelstein effect (EE) in the Tl-Pb layer.   } 
\label{fig1}
\end{center}
\end{figure}


Atomic-layer Tl-Pb compounds were prepared by a molecular beam epitaxy method, following the process established by some of the present authors \cite{matetskiy}. Pristine Tl/Si(111) - ($1 \times 1$) reconstruction was made by depositing one-monolayer Tl onto a Si(111) - ($7 \times 7$) surface at $\sim 300$ $^\circ {\rm C}$. Then, one-third monolayer of Pb was deposited onto the Tl/Si(111) surface at room temperature. The surface reconstruction was confirmed by monitoring reflection high-energy electron diffraction (RHEED) patterns during the growth process, as shown in Fig. \ref{fig1}(b). The RHEED pattern of the Tl-Pb compound differs from that of the parent Tl monolayer, and $\sqrt{3} \times \sqrt{3}$ periodicity appears. On top of the Tl-Pb compounds prepared on the Si substrates, a 60-nm-thick Cu capping layer was deposited with an in situ electron-beam evaporator. Since the spin diffusion length of Cu ($\sim 500$ nm \cite{yakata, bass}) is much longer than the thickness of the Cu layer, the loss of spin current in the Cu layer is negligible. The Cu/Tl-Pb bilayers were transferred to another vacuum chamber, and then a 20-nm-thick Py (Ni$_{81}$Fe$_{19}$) film was deposited onto them using electron beam evaporation to study the Edelstein effects in Py/Cu/Tl-Pb trilayer samples.    
\par


\begin{figure}[t]
\begin{center}
\includegraphics[width=8cm]{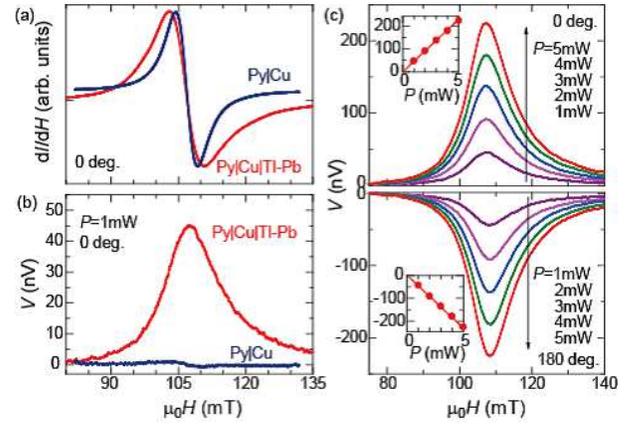}
\caption{(Color online.) (a) FMR derivative absorption spectra (${\rm d}I/{\rm d}H$) for Py/Cu/Tl-Pb and Py/Cu films. (b) Voltage spectra for Py/Cu/Tl-Pb and Py/Cu films. (c) Voltage spectra for Py/Cu/Tl-Pb at various input-microwave power ($P$) levels and different magnetic-field directions ($0$ deg. and $180$ deg.). The insets show the $P$ dependence of the voltage peak magnitudes at $0$ deg. and $180$ deg.   } 
\label{fig2}
\end{center}
\end{figure}

First, we investigated the inverse Edelstein effect induced by spin pumping into the Tl-Pb layer. The experiments were performed at room temperature using an electron spin resonance (ESR) spectrometer \cite{ando-jap}, as illustrated in Fig. \ref{fig1}(c). Py/Cu/Tl-Pb trilayer samples were placed in the center of the ESR cavity and  a $9.4$ GHz microwave was applied to the samples. When an external magnetic field applied along the film plane [$x$ axis in Fig. \ref{fig1}(c)] satisfies the FMR condition, the injection of spin currents into the Tl-Pb layer takes place via the spin pumping. DC voltages arising on the Tl-Pb layer along the $y$ axis in Fig. \ref{fig1}(c) around the FMR magnetic fields were measured using a nanovoltmeter. The length between the voltage electrodes is $2$ mm, and the sample width is $1$ mm.    
\par

Figure \ref{fig2}(a) shows an FMR derivative absorption spectrum, ${\rm d}I/ {\rm d}H$, for Py/Cu/Tl-Pb and Py/Cu samples. FMR of the Py layer is observed for both the samples at the magnetic field of $\sim 105$ mT. The FMR linewidth is clearly enhanced for the Py/Cu/Tl-Pb trilayer compared to the Py/Cu bilayer which does not include the Tl-Pb layer: peak-to-peak linewidth $\Delta H_{\rm p-p} =7.86$ mT for the Py/Cu/Tl-Pb trilayer and $\Delta H_{\rm p-p} =4.88$ mT for the Py/Cu bilayer. The enhancement of the FMR linewidth can be ascribed to the spin pumping; the transfer of spin currents from Py to the Tl-Pb layer effectively increases the damping of the Py magnetization precession. 
\par

Voltage spectra measured for Py/Cu/Tl-Pb and Py/Cu samples at around FMR magnetic fields are shown in Fig. \ref{fig2}(b). Here, the input microwave power is set at $1$ mW. For the Py/Cu/Tl-Pb trilayer, a positive voltage peak with the magnitude of $\sim 45$ nV is clearly observed at around the FMR magnetic-field, while no voltage peak is discerned for the Py/Cu bilayer. The line shape of the small voltage signal observed in the Py/Cu bilayer is similar to that of voltage induced by the anomalous Hall effect in the Py layer \cite{ando-jap}. At FMR, the Py magnetization precession generates spin currents. The resulting spin accumulation on the Tl-Pb layer is converted to a charge current by the inverse Edelstein effect. The Lorentz-type voltage peak observed in the Py/Cu/Tl-Pb trilayer is consistent with the inverse Edelstein effect \cite{sanchez, shiomi, deorani, jameli, Sn, yamamoto, HWang, smb6, karube, tsai, whan}.   
\par

When the direction of the magnetic field is reversed from the $+x$ ($0$ deg.) to $-x$ ($180$ deg.) direction, the voltage peaks in the Py/Cu/Tl-Pb sample change their sign as expected for the inverse Edelstein effect, as shown in Fig. \ref{fig2}(c). The magnitudes of the voltage peaks increase monotonically, as the input microwave power increases. As shown in the inset to Fig. \ref{fig2}(c), the microwave-power dependence is linear, which indicates that the observed spin pumping signals are in the linear regime. 
\par

The spin pumping signals in Fig. \ref{fig2} allow us to estimate the Edelstein length \cite{sanchez}, which is a measure of the conversion efficiency of the effect. From the FMR line widths shown in Fig. \ref{fig2}(a), the spin current injected into the Tl-Pb layer is estimated to be $1592$ A/m$^{2}$ using an established formula for spin pumping \cite{ando-jap, iguchi-jpsj}. Using the voltage magnitude shown in Fig. \ref{fig2}(b) and the sample resistance ($100$ ${\rm \Omega}$), the generated charge current is estimated to be $2.25 \times 10^{-7}$ A/m. Hence, the Edelstein length $\lambda_{EE}$ is obtained as the ratio of the charge to spin current: $0.14$ nm. This magnitude is comparable to the reported $\lambda_{EE}$ values for topological insulator Bi$_{2}$Se$_{3}$ ($0.21$ nm) \cite{deorani} and Sn-Bi$_{2}$Te$_{2}$Se ($0.27$ nm) \cite{yamamoto}, but 15 times smaller than that for $\alpha$-Sn ($2.1$ nm) \cite{Sn}.  
\par

\begin{figure}[t]
\begin{center}
\includegraphics[width=8cm]{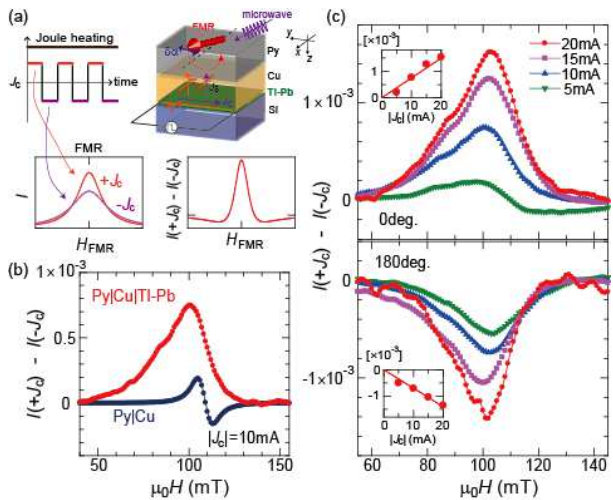}
\caption{(Color online.) (a) Schematic illustration of lock-in detection of the damping modulation, and predicted signal shapes of the experimental results. $I$ denotes the FMR absorption intensity. (b) The modulation signal of the FMR spectrum as a function of an external magnetic field for Py/Cu/Tl-Pb and Py/Cu. (c) The modulation signal at various electric-current ($J_{C}$) magnitudes and at different magnetic-field directions ($0$ deg. and $180$ deg.). The insets show the $|J_{C}|$ dependence of the modulation signals at $0$ deg. and $180$ deg.    } 
\label{fig3}
\end{center}
\end{figure}

To confirm the efficient spin-charge interconversion in the Tl-Pb compound, we next performed an experiment of the Edelstein effect in the same Py/Cu/Tl-Pb trilayer sample, as shown in Fig. \ref{fig3}. The experimental setup for the detection of the Edelstein effect is illustrated in Fig. \ref{fig1}(d). When an electric current $J_{C}$ is applied to the Tl-Pb layer along the $y$ axis, Tl-Pb exhibits the Edelstein effect, and thus the charge current is converted to spin accumulation in the Tl-Pb layer. The accumulated spin angular momentum is immediately transferred to the Py layer via a spin current propagating through the Cu layer along the $z$ axis. The magnetization-precession relaxation of the Py layer is modulated by the injected spin current, which can be detected as the modulation of the damping constant (line width) in the FMR spectra. 
\par

Several experiments of the damping modulation by spin transfer torque have been reported using the spin Hall effect in Py/Pt systems \cite{ando-prl, kasai}. In these experiments, DC charge currents were applied to Pt layers, which may cause additional damping modulation due to heating effects \cite{ando-prl}. To rule out extrinsic effects due to the heating effect, a lock-in detection is used in the present experiment, as shown in Fig. \ref{fig3}(a). When the direction of $J_{C}$ is reversed between $+y$ and $-y$ directions, the sign of the damping modulation induced by the Edelstein effect is reversed, whereas the heating effect does not change sign. Lock-in detection of the FMR spectra with the frequency of $J_{C}$ enables precise estimation of the damping modulation due to the Edelstein effect.  
\par

Figure \ref{fig3}(b) shows modulation of FMR absorption intensity ($I$) by the $J_{C}$ directions, $I(+J_{C})-I(-J_{C})$, measured for the Py/Cu/Tl-Pb trilayer and the Py/Cu bilayer in the ESR cavity. Signals are observed around the FMR magnetic field of the Py layer for both the samples, but clearly, the observed signal shapes are different between Py/Cu/Tl-Pb and Py/Cu. For the Py/Cu bilayer, the signal shape is dispersion-type, and its magnitude is small relative to the Py/Cu/Tl-Pb case. The dispersion-type signal may be explained by an alternate Oersted field caused by AC charge currents. In contrast, for the Py/Cu/Tl-Pb trilayer, the signal shape is a peak, similar to the predicted signal shape for the Edelstein effect [Fig. \ref{fig3}(a)].  
\par

The modulation signal is further studied by changing the $J_{c}$ magnitudes and the magnetic-field directions [$0$ deg. ($+x$ direction) and $180$ deg. ($-x$ direction)] in Fig. \ref{fig3}(c). As $|J_{C}|$ increases from $5$ mA to $20$ mA, the magnitudes of the peak signals monotonically increase. As shown in the insets to Fig. \ref{fig3}(c), the $|J_{C}|$ dependence is linear, which rules out the heating-induced extrinsic effects \cite{ando-prl}. Furthermore, when the direction of the magnetic field is reversed from the $+x$ to $-x$ direction, the peak signals keep their magnitude but change their sign. The results are consistent with the damping modulation due to the Edelstein effect.
\par

Let us quantitatively estimate the Edelstein length from the modulation signal shown in Fig. \ref{fig3}. Following the theoretical formulation developed for the damping modulation due to the spin Hall effect \cite{ando-prl}, the change in the effective damping constant induced by the Edelstein effect $\Delta \alpha_{EE}$ is given by
\begin{equation}
\Delta \alpha_{EE} = \frac{ \gamma \mu_{0} \hbar}{A_{F} \lambda_{EE} \omega M e} J_{C}.
\end{equation}  
Here, $\gamma$ is the gyromagnetic ratio, $A_{F} (= 2 \times 10^{-11}$ m$^2$) is the cross-sectional area of the Py layer, $\omega$ is the angular frequency of the microwave ($9.4$ GHz), $M$($=0.745$ T \cite{ando-jap}) is the Py magnetization, and the other symbols ($\mu_{0}$, $\hbar$, and $e$) have their usual meanings. The full width at half maximum of the modulation signal $\Delta H$ in Fig. \ref{fig3} corresponds to the difference in the damping modulation between $+J_{C}$ and $-J_{C}$ conditions: $(2 \omega/ \gamma) 2 \Delta \alpha_{EE}$. Hence, using $\Delta H \sim 20$ mT at $|J_{C}|=10$ mA in Fig. \ref{fig3}(b), we obtain $\lambda_{EE}=0.11$ nm from eq. (1). The similar $\lambda_{EE}$ value to that obtained by the inverse Edelstein effect (0.14 nm) supports the efficient Edelstein effects in the Tl-Pb compound.
\par

In summary, we studied spin-charge interconversion effects in Tl-Pb one-atom-layer material at room temperature. Two-dimensional Tl-Pb compounds which consist of one monolayer of Tl and one-third monolayer of Pb were grown on Si (111) substrates by a molecular beam epitaxy method. By spin pumping from ferromagnetic permalloy films into the Tl-Pb compounds, clear spin-charge conversion voltage is observed, which can be ascribed to the inverse Edelstein effect due to the strong Rashba-type spin-orbit interaction of the Tl-Pb layer. The measurement of the inverse effect of spin pumping, i.e. damping modulation by applied electric currents, confirms the high efficiency of the Edelstein effects in the Tl-Pb compounds. From the results of the direct and inverse Edelstein effects in the Py$\mid$Cu$\mid$Tl-Pb trilayers, the Edelstein length is estimated to be $0.11$-$0.14$ nm, which is comparable to that reported in the surfaces of topological insulators. The efficient Edelstein effects on Si substrates could be compatible with the semiconductor technology.
\par

This research was supported by JST ERATO ``Spin Quantum Rectification Project" (JPMJER1402), JSPS (KAKENHI No. 16H02108,  No. JP16H00983, No. 17H04806, No. 18H04215, and No. 18H04311 and the Core-to-Core program ``International research center for new-concept spintronics devices") and MEXT (Innovative Area ``Nano Spin Conversion Science" (No. 26103005)).
\par

%





\begin{thebibliography}{99}
\bibitem{weihan}W. Han, APL Materials {\bf 4}, 032401 (2016).
\bibitem{kush2006}M. S. Kushwaha, Phys. Rev. B {\bf 74}, 045304 (2006).
\bibitem{kush2007}M. S. Kushwaha, Phys. Rev. B {\bf 76}, 245315 (2007).
\bibitem{kush2008}M. S. Kushwaha, J. Appl. Phys. {\bf 104}, 083714 (2008).
\bibitem{Edelstein}V. M. Edelstein, Solid State Commun. {\bf 73}, 233-235 (1990).
\bibitem{sanchez}J. C. Rojas-S\'anchez, L. Vila, G. Desfonds, S. Gambarelli, J. P. Attan\'e, J. M. De Teresa, C. Mag\'en, and A. Fert, Nat. Commun. 4, 2944 (2013).
\bibitem{shiomi}Y. Shiomi, K. Nomura, Y. Kajiwara, K. Eto, M. Novak, Kouji Segawa, Yoichi Ando, and E. Saitoh, Phys. Rev. Lett. {\bf 113}, 196601 (2014).
\bibitem{deorani}P. Deorani, J. Son, K. Banerjee, N. Koirala, M. Brahlek, S. Oh, and H. Yang, Phys. Rev. B {\bf 90}, 094403 (2014).
\bibitem{jameli}M. Jamali, J. S. Lee, J. S. Jeong, F. Mahfouzi, Y. Lv, Z. Zhao, B. K. Nikolic, K. A. Mkhoyan, N. Samarth, and J.-P. Wang, Nano Lett. {\bf 15}, 7126 (2015).
\bibitem{Sn}J. C. Rojas-S\'anchez, S. Oyarzun, Y. Fu, A. Marty, C. Vergnaud, S. Gambarelli, L. Vila, M. Jamet, Y. Ohtsubo, A. Taleb-Ibrahimi, P. Le Fevre, F. Bertran, N. Reyren, J.-M. George, and A. Fert, Phys. Rev. Lett. 116, 096602 (2016).
\bibitem{yamamoto}K. T. Yamamoto, Y. Shiomi, Kouji Segawa, Yoichi Ando, and E. Saitoh, Phys. Rev. B 94, 024404 (2016).
\bibitem{HWang}Hailong Wang, James Kally, Joon Sue Lee, Tao Liu, Houchen Chang, Danielle Reifsnyder Hickey, K. Andre Mkhoyan, Mingzhong Wu, Anthony Richardella, and Nitin Samarth, Phys. Rev. Lett. {\bf 117}, 076601 (2016).
\bibitem{smb6}Qi Song, Jian Mi, Dan Zhao, Tang Su, Wei Yuan, Wenyu Xing, Yangyang Chen, Tianyu Wang, Tao Wu, Xian Hui Chen, X. C. Xie, Chi Zhang, Jing Shi, and Wei Han, Nature Commun. {\bf 7}, 13485 (2016).
\bibitem{karube}S. Karube, K. Kondou, and Y. Ohtani, Appl. Phys. Express {\bf 9}, 033001 (2016).
\bibitem{tsai}Hanshen Tsai, Shutaro Karube, Kouta Kondou, Naoya Yamaguchi, Fumiyuki Ishii,  Yoshichika Otani, Sci. Rep. {\bf 8}, 5564 (2018).
\bibitem{whan}Qi Song, Hongrui Zhang, Tang Su, Wei Yuan, Yangyang Chen, Wenyu Xing, Jing Shi, Jirong Sun, and Wei Han, Sci. Adv. {\bf 3}, e1602312 (2017).
\bibitem{ralph}A. R. Mellnik, J. S. Lee, A. Richardella, J. L. Grab, P. J. Mintun, M. H. Fischer, A. Vaezi, A. Manchon, E.-A. Kim, N. Samarth, and  D. C. Ralph, Nature (London) {\bf 511}, 449 (2014).
\bibitem{fan}Yabin Fan, Pramey Upadhyaya, Xufeng Kou, Murong Lang, So Takei, Zhenxing Wang, Jianshi Tang, Liang He, Li-Te Chang, Mohammad Montazeri, Guoqiang Yu, Wanjun Jiang, Tianxiao Nie, Robert N. Schwartz, Yaroslav Tserkovnyak, and  Kang L. Wang, Nat. Mater. {\bf 13}, 699 (2014).
\bibitem{lander}J. J. Lander, Surf. Sci. {\bf 1}, 125 (1964).
\bibitem{zhang}T. Zhang, P. Cheng, W. J. Li, Y. J. Sun, G. Wang, X. G. Zhu, K. He, L. Wang, X. Ma, and X. Chen {\it et al.}, Nat. Phys. {\bf 6}, 104 (2010).
\bibitem{qin}S. Qin, J. Kim, Q. Niu, and C. K. Shih, Science {\bf 324}, 1314 (2009).
\bibitem{uchihashi}T. Uchihashi, P. Mishra, M. Aono, and T. Nakayama, Phys. Rev. Lett. {\bf 107}, 207001 (2013).
\bibitem{yamada}M. Yamada, T. Hirahara, and S. Hasegawa, Phys. Rev. Lett. {\bf 110}, 237001 (2013).
\bibitem{uchihash2}T. Uchihashi, P. Mishra, and T. Nakayama, Nanoscale Res. Lett. {\bf 8}, 167 (2013).
\bibitem{yoshizawa}S. Yoshizawa, H. Kim, T. Kawakami, Y. Nagai, T. Nakayama, X. Hu, Y. Hasegawa, and T. Uchihashi, Phys. Rev. Lett. {\bf 113}, 247004 (2014).
\bibitem{noffsinger}J. Noffsiger and M. L. Cohen, Solid State Commun. {\bf 151}, 421 (2011).
\bibitem{zhao}W. Zhao, Q. Wang, M. Liu, W. Zhang, Y. Wang, M. Chen, Y. Guo, K. He, X. Chen, and Y. Wang {\it et al.}, Solid State Commun. {\bf 165}, 59 (2013).
\bibitem{brun}C. Brun, T. Cren, V. Cherkez, F. Debontridder, S. Pons, D. Fokin, M. C. Tringides, S. Bozhko, L. B. Ioffe, and B. L. Altshuler, {\it et al.} Nat. Phys. {\bf 10}, 444 (2014).
\bibitem{matetskiy}A. V. Matetskiy, S. Ichinokura, L. V. Bondarenko, A. Y. Tupchaya, D. V. Gruznev, A. V. Zotov, A.?A. Saranin, R. Hobara, A. Takayama, and S. Hasegawa, Phys. Rev. Lett. {\bf 115}, 147003 (2015).
\bibitem{ast}Christian R. Ast, Jurgen Henk, Arthur Ernst, Luca Moreschini, Mihaela C. Falub, Daniela Pacil\'e, Patrick Bruno, Klaus Kern, and Marco Grioni, Phys. Rev. Lett. {\bf 98}, 186807 (2007).
\bibitem{gierz}I. Gierz, T. Suzuki, E. Frantzeskakis, S. Pons, S. Ostanin, A. Ernst, J. Henk, M. Grioni, K. Kern, and C. R. Ast, Phys. Rev. Lett. {\bf 103}, 046803 (2009).
\bibitem{sakamoto}Kazuyuki Sakamoto, Haruya Kakuta, Katsuaki Sugawara, Koji Miyamoto, Akio Kimura, Takuya Kuzumaki, Nobuo Ueno, Emilia Annese, Jun Fujii, Ayaka Kodama, Tatsuya Shishidou, Hirofumi Namatame, Masaki Taniguchi, Takafumi Sato, Takashi Takahashi, and Tamio Oguchi, Phys. Rev. Lett. {\bf 103}, 156801 (2009).
\bibitem{frantzeskakis}E. Frantzeskakis, S. Pons, and M. Grioni, Phys. Rev. B {\bf 82}, 085440 (2010).
\bibitem{sakamoto2}Kazuyuki Sakamoto, Tatsuki Oda, Akio Kimura, Koji Miyamoto, Masahito Tsujikawa, Ayako Imai, Nobuo Ueno, Hirofumi Namatame, Masaki Taniguchi, P. E. J. Eriksson, and R. I. G. Uhrberg, Phys. Rev. Lett. {\bf 102}, 096805 (2009).
\bibitem{azpiroz}J. Iba\~nez-Azpiroz, A. Eiguren and A. Bergara, Phys. Rev. B {\bf 84}, 125435 (2011).
\bibitem{sakamoto3}Kazuyuki Sakamoto, Tae-Hwan Kim, Takuya Kuzumaki, Beate Muller, Yuta Yamamoto, Minoru Ohtaka, Jacek R. Osiecki, Koji Miyamoto, Yasuo Takeichi, Ayumi Harasawa, Sebastian D. Stolwijk, Anke B. Schmidt, Jun Fujii, R. I. G. Uhrberg, Markus Donath, Han Woong Yeom, and Tatsuki Oda, Nat. Commun. {\bf 4}, 2073 (2013).
\bibitem{stolwijk}S. D. Stolwijk, A. B. Schmidt, M. Donath, K. Sakamoto, and P. Kr\"uger, Phys. Rev. Lett. {\bf 111}, 176402 (2013).
\bibitem{stolwijk2}S. D. Stolwijk, K. Sakamoto, A. B. Schmidt, P. Kr\"uger, and M. Donath, Phys. Rev. B {\bf 90}, 161109 (2014).
\bibitem{yakata}S. Yakata, Y. Ando, T. Miyazaki, and S. Mizukami, Jpn. J. Appl. Phys. {\bf 45}, 3892-3895 (2006).
\bibitem{bass}J. Bass and W. P. Pratt Jr., J. Phys.: Condens. Matter {\bf 19} 183201 (2007).
\bibitem{ando-jap}K. Ando, S. Takahashi, J. Ieda, Y. Kajiwara, H. Nakayama, T. Yoshino, K. Harii, Y. Fujikawa, M. Matsuo, S. Maekawa, and E. Saitoh, J. Appl. Phys. {\bf 109}, 103913 (2011).
\bibitem{iguchi-jpsj}R. Iguchi and E. Saitoh, J. Phys. Soc. Jpn. {\bf 86}, 011003 (2017).
\bibitem{ando-prl}K. Ando, S. Takahashi, K. Harii, K. Sasage, J. Ieda, S. Maekawa, and E. Saitoh, Phys. Rev. Lett. {\bf 101}, 036601 (2008).
\bibitem{kasai}S. Kasai, K. Kondou, H. Sukegawa, S. Mitani, K. Tsukagoshi, and Y. Ohtani, Appl. Phys. Lett. {\bf 104}, 092408 (2014).
\end{thebibliography}
\end{document}